\documentclass[prb,aps,twocolumn,showpacs,superscriptaddress]{revtex4-1}
\usepackage{hyperref}
\hypersetup{colorlinks, citecolor=blue, filecolor= black, linkcolor= black, urlcolor= blue}
\usepackage{graphicx}% Include figure files
\graphicspath{ {phase-diag/} {regimes/} }
\usepackage{dcolumn}% Align table columns on decimal point
\usepackage{bm}% bold math
\usepackage{float}
\usepackage{bm}
\usepackage{epsfig}
\usepackage{latexsym}
\usepackage{mathtools}
\usepackage{amsfonts}
\usepackage{color}
\usepackage{array}
\usepackage{enumitem}
\usepackage{hyperref}
\hypersetup{
    colorlinks=true,
    linkcolor=blue,
    filecolor=magenta,      
    urlcolor=cyan,
}
\usepackage{lipsum}
\usepackage{amssymb, amsmath}

\DeclarePairedDelimiter\ket{\lvert}{\rangle}
\DeclarePairedDelimiterX\braket[2]{\langle}{\rangle}{#1 \delimsize\vert #2}

\renewcommand{\>}{\rangle}
\renewcommand{\(}{\left(}
\renewcommand{\)}{\right)}

\renewcommand{\d}{\partial}

\begin{document}
	\preprint{Preparing}

	%=====================  Title   ========================%
	\title{Interlayer coherence in Superconductor Bilayers}
	
	%=====================  Authors   =====================%
	\author{Igor V. Blinov}
	\email{blinov@utexas.edu}
	\affiliation{
		Department of Physics, The University of Texas at Austin, Austin, Texas 78712, USA}
	\author{Allan H. MacDonald}
	\affiliation{
		Department of Physics, The University of Texas at Austin, Austin, Texas 78712, USA}

	\date{June 30, 2022}
	
	%=====================  Abstract   =====================%
	\begin{abstract}
		\noindent
			We investigate the possibility and implications of interlayer coherence in electrically isolated superconducting bilayers.
			We find that in a mean-field approximation bilayers can have superconducting and 
			excitonic order simultaneously if repulsive interactions between layers are sufficiently strong.
			The excitonic order implies interlayer phase coherence, and can be studied in a representation of symmetric and antisymmetric bilayer states.  When both orders are present we find several solutions of the mean-field equations with different values of the the symmetric and antisymmetric state pair amplitudes. 
		The mixed state necessarily has non-zero pair amplitudes for electrons in 
			different layers in spite of the repulsive interlayer interactions, and these 
			are responsible for spatially indirect Andreev reflection processes in which 
			an incoming electron in one layer can be reflected as a hole in the opposite layer.  
			We evaluate layer diagonal and off-diagonal current-voltage relationships that can be used to identify this state experimentally.
				\end{abstract}

	%\keywords{}
	\maketitle
	\newpage
	%\begin{multicols}
	
	%=====================  Main Context

\section{Introduction}
Superconductivity is a well-explored yet still puzzling phenomena of collective behavior of electrons. 
The microscopic BCS \cite{bardeen1957theory,bardeen1957microscopic} theory explains it in terms of bound (Cooper) electron pairs due to effective attractive interactions. Condensation of Cooper pairs gives rise to a non-Fermi-liquid state with spontaneously broken $U(1)$ symmetry characterized by a homogenous complex order parameter $\Delta$.  One signature  of superconductivity is 
a non-linear junction resistance with a normal metal.  
%due to processes in which electrons incident from the normal 
%metal are reflected as holes, launching Cooper pairs inside the superconductor to conserve charge.
At subgap values of the bias voltage on the interface, the only way for a single electron to penetrate inside a 
superconductor is to form a Cooper pair. Therefore, by charge conservation, transmission through the interface should be accompanied by reflection of a hole, a process known as Andreev \cite{andreev1965thermal} reflection.  Studies of 
Andreev reflection can be used to probe \cite{bradleySuperfluidAndreev, beenakkerColloquim} a state of interest and continue to gather both theoretical 
\cite{de1995andreev,diez2012andreev,beenakker2006specular,nilsson2008splitting,stanescu2011majorana,kundu2013transport,mazin1999degreeofspinpol,mazin2001probing} and experimental \cite{sasaki2011topological,kastalsky1991observation,pothier1994flux,park2005andreev} attention.  

In recent years experimentalists have uncovered examples of both electron-electron (Cooper) and 
electron-hole (excitonic) pairing in atomically thin two-dimensional materials\cite{xi2016ising, li2018nontrivial,hsu2020inversion,eisensteinQHE, simonet2017anomalous,tutucEnhancement}.  
In general Cooper pairing is driven by 
attractive effective interactions, and electron-hole pair formation by repulsive interactions between electrons
in different bands.  Exciton condensates \cite{keldysh1968collective,lozovik1975feasibility, hanamura1977condensation, littlewood2004models,perali2013high,fogler2014high} 
are coherent states of electrons and holes bound into pairs by the Coulomb interaction, and are
described by a mean-field theory that is identical
to BCS theory apart from a particle-hole transformation.  
In bulk materials the concept of exciton condensation is partly ambiguous since the ordered states are difficult to distinguish from density-wave or nematic states \cite{fradkin2010nematic}.
Although exciton condensates were predicted theoretically more than 50 years ago, 
their experimental identification in bulk materials has suffered as a consequence,
although progress has been achieved recently\cite{kogar2017signatures,mazuz2019dynamical, high2012condensation}.
Experimental advances with two-dimensional materials have solved the ambiguity problem 
by making it possible to prepare devices with strong interactions between subsystems 
located in different layers that separately have nearly perfectly conserved particle number.
Interlayer coherence is a related yet slightly different phenomena. It corresponds to the same order parameter and also breaks the independent gauge invariance of the individual layers:
{\it i.e.} when the exciton order parameter $x$ is nonzero, we no longer can perform gauge transformation in two layers independently. However, interlayer coherence, unlike the exciton condensation, does not open a gap in the system, and bears more similarity to pseudospin magnetism.
There are still only a few examples of convincing experimental evidence for the excitonic condensation and interlayer coherence in systems without 
a strong external magnetic field, but this situation seems likely to change.

Recently, superconductivity has been observed in magic angle twisted bilayer 
graphene (tBG) \cite{bistritzer2011moire, cao2018unconventional}, a two-dimensional system with 
a moir\'e superlattice that yields extremely flat conduction and valence bands.  
The strongest superconductivity appears when the 
valence band is doped away from half filling.  Although several models \cite{wu2018theory, lian2019twisted, kozii2019nematic, chichinadze2020nematic, khalaf2020charged} for its superconductivity have been proposed, the mechanism remains unknown.
%\iffalse As a reference, we use the model with phonon-mediated superconductivity \cite{wu2018theory}. It predicts the strength of the %attractive electron-electron $\lambda_s$ interaction to be around $50 \ meV nm^2$  \fi
Twisted bilayers are the simplest examples of a rich variety of graphene multilayer moir\'e superlattice systems that 
have received experimental attention recently \cite{cao2018correlated,lu2019superconductors,yankowitz2019tuning,park2021tunable,rickhaus2020density}. The motivation for this paper is to 
consider the possibility of realizing, in this flexible family of strongly-correlated electron systems, an exotic state in which interlayer coherence and superconductivity 
occur simultaneously.  Our target system is two twisted bilayers separated by 
a hexagonal boron nitride tunnel barrier.  The superconductivity of the individual 
twisted bilayers is already established.  Here we ask the following questions: i) can interlayer 
coherence occur in principle between two-dimensional electron systems that are 
superconducting ii) if so, how would such a state be most unambiguously detected?  So far 
research on the coexistence of two phases, or, broadly speaking, multi-component coupled condensates, has been scarce \cite{sager2020potential,conti2017multicomponent,vargas2020crossband} partially because of the lack of a physical system with a potential to have both. In this paper, we study a possibility for a system to have both interlayer coherence (magnetism in layer-pseudospin label) and superconductivity within each layer: a state in which two superconducting states within each layer are coupled through the spontaneously established interlayer coherence.
In section II  we describe the model and study and relationships between its order parameters.
In section III we describe the phases we have identified and phase diagrams as a function of the strength of the 
interlayer repuslive interaction strength $g_x$ and the intralayer attractive interaction strength
$g_s$.  In the section IV we propose an Andreev drag measurement which can be used to
identify states with both types of order, taking advantage \cite{su2008make} of the possibility of
separate contacting to individual layers.
This method of probing bilayer exciton condensates has been 
very successfully exploited in quantum Hall excitonic superfluids \cite{liu2017quantum,eisenstein2014exciton,yoon2010interlayer,li2017excitonic,li2019pairing}
Finally, in section V we discuss prospects for observing these 
states and the importance of deviations from the highly symmetric point that we considered in our explicit calculations.
\iffalse
There are two main scenarios for spatially indirect exciton condensation. 
In the first, a system is symmetric under the combination of layer interchange and
an electron-hole transformation. In this case, an electron-hole transformation in one
of the layers maps exciton condensates to superconductors that pair electrons in different layers and 
occurs for arbitrarily weak interactions.
Consequently, charge transmission at low voltage occurs via a process 
similar to Andreev reflection: to transfer an electron through an interface in one of the layers, 
an electron should be reflected from the bottom layer.

In a scenario with two identical layers with the same Fermi level, exciton condensation is similar to itinerant electron
ferromagnetism with the spin degree-of-freedom replaced by a layer degree of freedom.  The phase will appear if the interaction $\lambda_x$ between the electrons and holes in two different layers is strong enough $g_x\equiv \lambda_x \nu(\epsilon_F)\geq 1$, where $\nu(\epsilon_F)$ is the density of states on the Fermi level. 
 In this case, the eigenstates of the mean-field Hamiltonian have layer spinors that are coherent combinations of the 
 individual layers, for example symmetric and antisymmetric.  
 When viewed in this basis, transmission through the interface will be similar to free electron reflection from a 
 potential.
  \fi

\section{Model}
We consider the simplest possible model that can have both interlayer coherence and s-wave superconductivity. 
The model assumes that both the attractive intralayer and repulsive interlayer 
interactions are momentum-independent, includes a layer degree of freedom ($l=t,b$), 
and assumes either valley or spin singlet superconductivity.
The  Hamiltonian
\begin{multline}\label{model:full-hamiltonian}
	H
	=
	\sum_{pl\sigma}\xi_{pl}c^{\dagger}_{l\sigma}(p)c_{l\sigma}(p)
	\\+
	\frac{\lambda_s}{S}\sum_{pp'q} 
	c^{\dagger}_{l\uparrow}(p+q)
	c^{\dagger}_{l\downarrow}(p'-q)
	c_{l\downarrow}(p')
	c_{l\uparrow}(p)
	\\+
	\frac{\lambda_x}{S}
	\sum_{pp'q} 
	c^{\dagger}_{t\sigma}(p+q)
	c^{\dagger}_{b\sigma'}(p'-q)
	c_{b\sigma'}(p')
	c_{t\sigma}(p),
\end{multline}
where $c^\dagger_{l\sigma}$ $(c_{l\sigma})$ creates (annihilates) a 
fermion in layer $l=t,b$ with pseudospin $\sigma=\uparrow,\downarrow$, and $S$ is the area of the sample, $\xi_{pl\sigma}=\epsilon(p)_{l\sigma}-\mu_{l}$ where $\epsilon(p)_{l\sigma}$ is a dispersion relation and $\mu_{l}$ is the Fermi energy within the layer.
In this work, we will focus on the case of layer-independent bands,
$\xi_{pl}\equiv \xi_p$, and reserve discussion of deviations from this symmetric limit to the end of the paper.
(As we explain in Section V coexistence does not occur for $\xi_{pt}\equiv -\xi_{pb}$.)
The pseudospin label refers to real spin in the case of spin-singlet superconductors and to
valley in the case of valley-singlet superconductors.  Below we refer to this degree of freedom as spin for simplicity.  
  \begin{figure}[htb!]
	\includegraphics[width=\columnwidth]{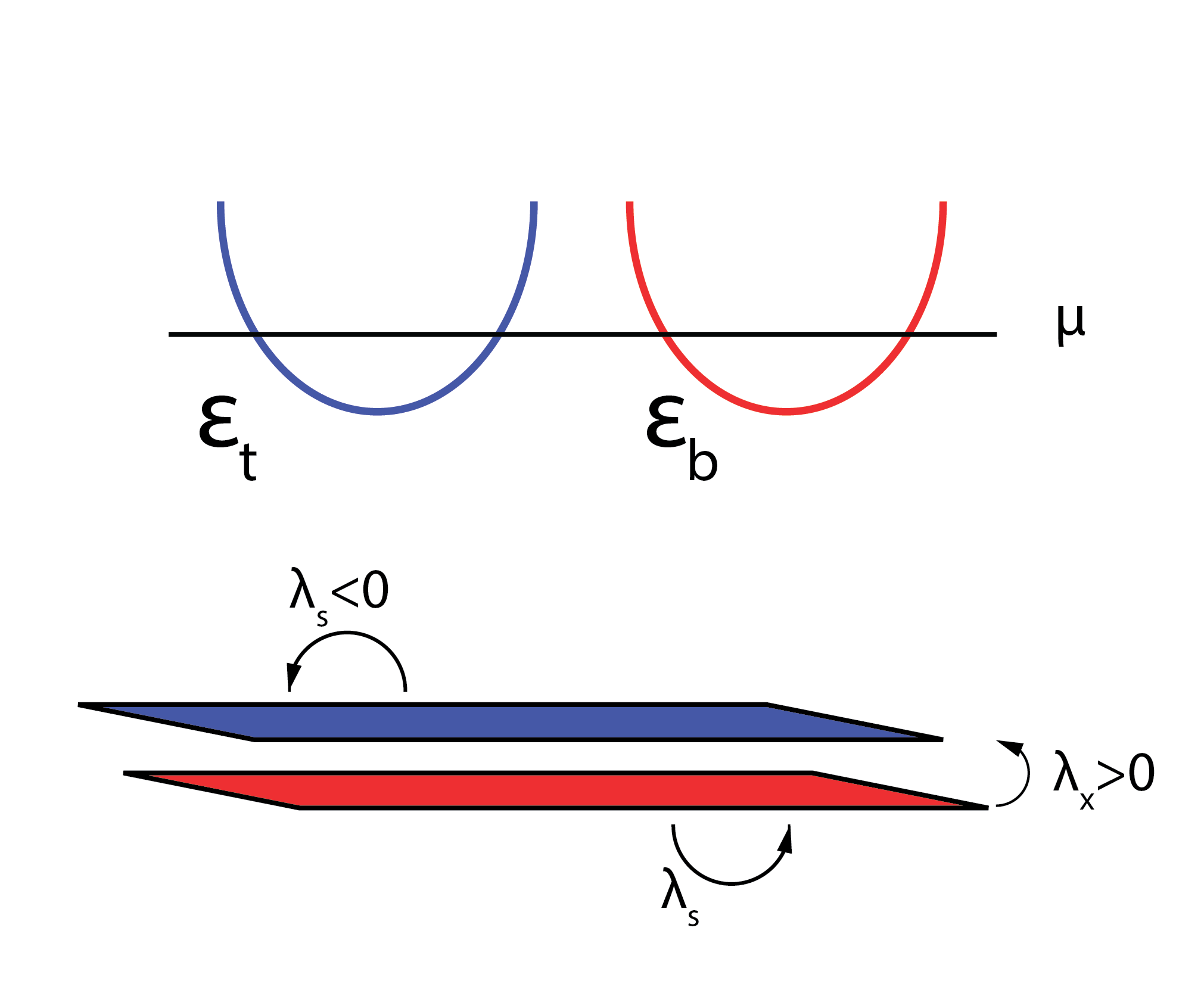}
	\caption{Sketch of the band alignment and the configuration of the layers. We take energy dispersion to be the same in both layers as well as the electron filling. Interactions within the layer ($\lambda_s<0$) are attractive while between the layers ($\lambda_x>0$)  is repulsive.
	We talk about deviations from the perfectly symmetric scenario in the discussion section. }
	\label{fig:model:bands}
\end{figure}
Below we characterize the ordered states by their mean-field Hamiltonians,
and allow an exciton order parameter for each spin species,
\begin{equation}
	x_\sigma=\frac{\lambda_x}{S}\sum_p\<c_{t\sigma}(p)c_{b\sigma}^\dagger(p)\>,
\end{equation}
and superconducting order parameters both within and between layers,
\begin{equation}
	\Delta^{ll'}_{\sigma\sigma'}=-\frac{\lambda_{ll'}}{S}\sum_p\<c_{l\sigma}(p)c_{l'\sigma'}(-p)\>,
\end{equation}
where $\lambda_{tt}=\lambda_{bb}=\lambda_s$ and $\lambda_{tb}=\lambda_{bt}=\lambda_x$.
The ordered state is therefore 
characterized by six complex self-energies that vanish if no symmetries are broken. 

In a single-band superconductors, the pair potential can be chosen to be real because of global gauge invariance.
It follows that one real number, the gap, fully characterizes the superconducting phase.  
In the present two-band system, we start with four complex pair potentials.
If we also allow excitonic particle-hole pairing for both spins, the ordered state is characterized by 
12 real numbers.  Since only three phases can be chosen at will by exploiting the conservation of the number of particles of particles of 
each spin in each layer, the energy can depend not only on absolute values of the order parameters, 
but also on their phases, or, more specifically, on gauge invariant combinations of such phases.

 In the absence of interlayer superconducting coupling, there is only one gauge invariant combination of the phases: $\psi_t-\psi_b-\psi_{\uparrow}-\psi_{\downarrow}$, where $\psi_{\sigma=\uparrow,\downarrow}$ is the phase of the corresponding exciton pair potential and $\psi_{l=t,b}$ is the phase of the intralayer superconducting pairing potential. This combination changes sign  under layer flip. If we assume that the ground state preserves spin and layer invariance, 
 we see that 
 \begin{equation}
     \psi_t-\psi_b-\psi_{\uparrow}-\psi_{\downarrow}=\pi n,
 \end{equation} where $n$ is an integer. 
 In the gauge $\psi_b=0$, $\psi_{\uparrow}=0$, $\psi_{\downarrow}=0$, an even $n$ will correspond to a state with two superconducting gaps having the same sign, while an odd $n$ will correspond to a state with superconducting gap that changes sign under the layer flip.

To extend this analysis to the case with nonzero interlayer pairing, 
we identify four gauge invariant quantities that do not transform to themselves under layer or spin inversion:
$x_\uparrow \Delta^{bt}_{\uparrow\downarrow} {\Delta^{tt}_{\uparrow\downarrow}}^*$, $x_\downarrow \Delta^{bt}_{\downarrow\uparrow}{\Delta^{tt}_{\downarrow\uparrow}}^*$, $x^*_\uparrow \Delta^{tb}_{\uparrow\downarrow}{\Delta^{bb}_{\uparrow\downarrow}}^*$, and $x^*_\downarrow \Delta^{tb}_{\downarrow\uparrow}{\Delta^{bb}_{\downarrow\uparrow}}^*$
and require that their values do not change under these transformations. \footnote{One way to represent these quantities is to imagine space with 4 points corresponding to each kind of fermion. Then any gauge invariant quantity will correspond to a closed trajectory between these points.}
This condition restricts the absolute values of all self-energies 
$|\Delta^t|=|\Delta^b|\equiv\Delta_d$, $|x_{\uparrow}|= |x_{\downarrow}|\equiv x$, $|\Delta^{tb}_{\uparrow\downarrow}|=|\Delta^{bt}_{\uparrow\downarrow}|\equiv \Delta_i$ as well as two gauge invariant combinations of phases 
\begin{align}\label{phase-cond:1}
\psi_1-\psi_2=\psi_\downarrow-\psi_\uparrow-2\pi m,\\\label{phase-cond:2} \psi_t-\psi_b=\psi_\uparrow+\psi_\downarrow-2\pi k,
\end{align}
where $\psi_1$ is the phase of $\Delta^{bt}_{\uparrow\downarrow}$ and $\psi_2$ is the phase of  $\Delta^{tb}_{\uparrow\downarrow}$, $k$ and $m$ are integers.
($d$, and $i$ in $\Delta_d$ and $\Delta_i$ are 
intended to suggest {\it direct} and {\it indirect} pairing.) 

The system has one more independent gauge invariant combination of phases $\psi_t+\psi_b-\psi_1-\psi_2\equiv2\psi_+$. Unlike those we considered previously, this combination transforms to itself under 
layer or spin flip, and therefore should be determined by energy minimization. 

Using global gauge invariance related to conserved 
particle numbers for each spin and valley $\psi_1$, $\psi_\uparrow$ and  $\psi_\downarrow$ can be chosen to be 0.  Then because of \eqref{phase-cond:1} $\psi_2=2\pi m$ and $\psi_-\equiv (\psi_t-\psi_b)/2=\pi k$. In what follows, we use this gauge.
This leaves us with 
mean-field Hamiltonians that depend only on four parameters.  There are three absolute values: $\Delta_d$ (direct superconducting gap, particle-particle pairing within a layer), $\Delta_i$ (indirect superconducting gap, particle-particle pairing between the layers), $x$ (electron-hole coherence between the two layers) and
a single free phase $(\psi_+\equiv(\psi_t+\psi_b)/2)$, 
whose values can be determined by solving self-consistent field equations.

\section{Phase Diagram}
We saw previously that layer and spin symmetries restrict the value of $\psi_-$, but not 
$\psi_+\equiv(\psi_t+\psi_b)/2$. In this section, we allow both phases to take any value to demonstrate explicitly that energies are minimized when spin/layer symmetries are respected: the four distinct allowed
values of $\psi_-$ ($0$, $\pi/2$, $3\pi/2$ $\pi$ if $\Delta_i=0$ and $0$ $\pi$ if $\Delta_i\neq 0$) follow from the self-consistency equations.
The mean-field version of \eqref{model:full-hamiltonian} can be written as 
\begin{equation}\label{phases:ham-matrix}
	H_{mf}=\sum_p \Psi^\dagger(p)
\mathcal{H}(p)
	\Psi(p),
\end{equation}
where $\Psi^\dagger(p)=\(c^\dagger_{t\uparrow},c_{t\downarrow},c^\dagger_{b\uparrow},c_{b\downarrow}\)$ is a vector in an extended Nambu space and the Hamiltonian matrix is
\begin{widetext} 
\begin{equation}
  \mathcal{H}(p)=	\begin{pmatrix}
		\xi_p & \Delta_d e^{i(\psi_++\psi_-)} & x & \Delta_i \\
		\Delta_d e^{-i(\psi_++\psi_-)}&-\xi_p & \Delta_i & -x\\
		x & \Delta_i & \xi_p &\Delta_d e^{i(\psi_+-\psi_-)}\\
		\Delta_i & -x & \Delta_d e^{-i(\psi_+-\psi_-)} &-\xi_p
	\end{pmatrix}
\end{equation}.
\end{widetext} 
The negative eigenvalues of the matrix have the form $\epsilon_{\pm}=-\(\eta^2\pm 2\alpha^2\)^{1/2}$, where $\eta^2=\xi^2+x^2+\Delta_i^2+\Delta^2_d$ and $\alpha^4=x^2 \Delta^2_d \sin^2(\psi_-)+2\Delta_d \Delta_i x\xi \cos(\psi_+)\cos(\psi_-)+\Delta_i^2\Delta_d^2\cos^2(\psi_+)+(x\xi)^2$.
Stable phases of the system are determined by minimization of the energy density.
At energy extrema,  
\begin{equation} \label{phases:total-energy}
	E=\frac{2\Delta^2_d}{|\lambda_s|}-\frac{2\Delta_i^2}{\lambda_x}+\frac{2 x^2}{\lambda_x}+\sum_b\int d\xi \nu(\xi )\epsilon_{b}n_b(\xi),
\end{equation}
where the sum is over quasiparticle bands $b$, $\nu(\xi)$ is the quasiparticle density-of-states per spin and per layer,
and $n_b(\xi)$ is the Fermi occupation factor.
At zero temperature, only quasiparticle states with negative energies are filled. 
%(Note that because the quasiparticle density-of-states depends on the order parameters, 
%the presence of a term proportional to $-|\Delta_i|^2$
%does not imply that increasing $\Delta_i$ will decrease energy without bound.)
\iffalse
 \begin{figure}[htb!]
	\includegraphics[width=\columnwidth]{Figures/energy_contour_vs_phi_fixed_orders_cust_cmap.png}
	\caption{Energy density \eqref{phases:total-energy} as a function of $\psi_-$ (x-axis) and $\psi_+$ (y-axis). Absolute values of all three order parameters fixed at some realistic values (on the main figure $\mu=0.1\Lambda$ $x=\mu$, $\Delta_d=0.01\Lambda$, $\Delta_i=\Delta_d/3$, and the density of states is the same as on Fig.\ref{fig:parallel-phase:underdoped} B), on the inset $\Delta_i=0$). The highest energy corresponds to red, the lowest energy corresponds to blue. We see that $\Delta_i$ lifts the energy degeneracy in $\psi_+$. Note that all minima are connected by $\psi_l\to \psi_l+2\pi n, \psi_l'\to\psi_l'+2\pi m$ for some integer $m$ and $n$ and therefore correspond to the same state. }
	\label{fig:parallel-phase:energy-vs-phi}
\end{figure}
\fi
Extrema of the energy are also extrema of Eq.~\eqref{phases:total-energy}, which we vary first 
with respect to the phases $\psi_-$ and $\psi_+$.
We conclude that energy minima occur at extrema of $\alpha^2$, and that these 
occur (independent of $\xi_p$) 
when 
\begin{equation}\label{phase-diagram:parallel-condition}
	\sin\(\psi_+\)=0 \text{ and } 	\sin\(\psi_-\)=0 
	\end{equation}
	or 
	\begin{equation}\label{phase-diagram:antiparallel-condition}
	\cos\(\psi_+\)=0 \text{ and } 	\cos\(\psi_-\)=0.
\end{equation}
The former condition is consistent with the conditions we derived for a mirror- and spin-symmetric state with $\Delta_i\neq0$. 
We will call it a parallel phase.  The latter condition defines the antiparallel phase \cite{phase-condition-comment}. 
We will see that the latter is consistent only with $\Delta_i=0$. 
Note that each condition in fact corresponds to two different states. Namely, \eqref{phase-diagram:parallel-condition} can be satisfied with $(\psi_t,\psi_b)=(\pi,\pi)$, and $(0,0)$, that typically have different energies 
when interlayer superconducting order is present.  Now we consider the antiparallel and parallel phases separately.

\subsection{Antiparallel phase}
We consider first extrema in which the phases $\psi_t$ and $\psi_b$ differ by $\pi$.
. %On Fig.\ref{fig:parallel-phase:energy-vs-phi} this combination of phases corresponds to an energy maximum for a particular values of order parameters.%
The main conclusion of this subsection is that for any sufficiently smooth density of states, 
the energy of the antiparallel phase is always higher than that of a pure superconducting phase.  
In other words this solution of the mean-field equations corresponds to a local energy maximum not 
an energy minimum, and can therefore be discarded.

  \begin{figure*}[t]
	\includegraphics[width=18cm,height=5.5 cm]
	{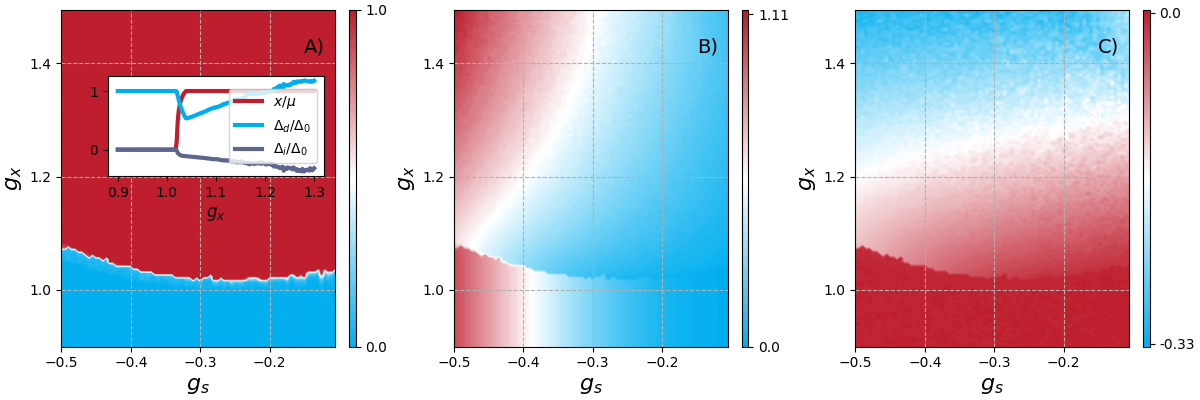}
	\caption{Order parameter dependence on dimensionless interaction constants $g_s=\nu \lambda_s$, 
	and $g_x=\nu \lambda_x$ in the parallel phase,
	where $\nu$ is the parabolic band density-of-states.
	The normalized exciton self-energy $x/\mu$ is plotted in panel A, where $\mu$ is the Fermi energy.  
	At small $|g_s|$ there is a phase transition at $g_x=1$, which is first order in our model, to a large 
	$g_x$ state with $x \ne 0$.  The normalized intralayer superconducting self-energy $\Delta_d/\Delta_0$, where $\Delta_0$ its maximal value in the pure superconducting phase, plotted in 
	panel B is always non zero.  
	The normalized interlayer superconducting self-energy $\Delta_i/\Delta_0$ is plotted 
	in panel C and acquires a finite value only when 
	$x \ne 0$.  At the transition the purely intralayer superconducting phase 
	to the mixed phase, $\Delta_d$ (B) decreases. For large values of $|g_s|$ the excitonic transition is pushed to larger values of $g_x$. The inset in panel A shows the three self-energies as a function of $g_x$ at a fixed $g_s=-0.4$. These results were obtained for cutoff $\Lambda=10 \ meV$, $\epsilon_F=3 \ meV$, and a parabolic band model with a constant density of states. To treat both $f=0$ and $f=1$ minima on the equal footing, we allowed the interlayer superconducting gap $\Delta_i$ to take negative values. Self-energies are in units of meV. We minimized energy as a function of the Bogolyubov angles using the conjugate graduate method.}
	\label{fig:parallel-phase:underdoped}
\end{figure*}
In the antiparallel phase, variation of the energy density in Eq.~\ref{phases:total-energy} with respect to ${\Delta_i}$ combines with Eq.~\eqref{phase-diagram:antiparallel-condition} to yield the self-consistency equation 
\begin{equation}
	\Delta_i
	=
	-\frac{\lambda_x \Delta_i}{4}\int\nu(\xi)\(\frac{1}{\epsilon_{+}}+\frac{1}{\epsilon_{-}}\).
\end{equation}
This equation clearly cannot be satisfied at any $\lambda_x>0$.  We conclude that 
$\Delta_i=0$ in the antiparallel phase. 
The quasiparticle band energies are therefore
\begin{equation}\label{antiparallel:band-energy}
	\pm \epsilon_{\pm}=\pm |\sqrt{\xi^2+\Delta_d^2}\pm x|\\
\end{equation} 
The form of the band energies allows us to identify
the antiparallel phase as an exciton condensate (spontaneous interlayer phase coherent state)
formed on top of superconductors within each of the layers.
With ${\Delta_i}$ eliminated, the system of gap equations reduces to:
\begin{equation}\label{phase-diagram:antiparallel:x}
	x=
	\lambda_x
	\mathcal{N}
	(\sqrt{x^2-\Delta_d^2} \,),
\end{equation}
\begin{multline}\label{phase-diagram:antiparallel:delta}
	\Delta_d=-\frac{\lambda_s \Delta_d}{2}
	\int  \frac{\nu(\xi)}{\sqrt{\xi^2+\Delta_d^2}}+\\
	\frac{\lambda_s \Delta_d}{2}\int^{\sqrt{x^2-\Delta_d^2}}_{-\sqrt{x^2-\Delta_d^2}}\frac{\nu(\xi)}{\sqrt{\xi^2+\Delta_d^2}}.
\end{multline}
That is to say that the exchange self-energy is contributed by wavevectors that have different 
occupation numbers for the two orthogonal quasiparticle layer spinors. 
In Eq.~\eqref{phase-diagram:antiparallel:x}  $\mathcal{N}(\epsilon) \equiv 1/2\int_{-\epsilon}^{\epsilon} d\xi \ \nu(\xi)$.

\iffalse
A transition between two superconductors and the mixed phase becomes possible at the smallest value of the interlayer interaction strength $\lambda_x$ where the gap equations can be satisfied.  From equation \eqref{phase-diagram:antiparallel:x} it follows that at the transition point the value $\frac{x}{\mathcal{N}(\sqrt{x^2-\Delta_d^2})}$,  has the minimum possible value.  For any given $x$ and any continuous density of state it implies that $\Delta_d=0$, so that at a transition point the superconducting gap in each layer should close. It is clear then from \eqref{phase-diagram:antiparallel:delta} that at the same point $x$ should be nonzero to ensure the gap closing. We conclude then that the possible transition between two phases is of the first order.
\fi
A transition between pure superconductivity and an antiparallel 
phase would occur at zero temperature if the latter phase were lower in energy.  
We find that 
\begin{multline}
\label{eq:energydiff}
	E_{m}-E_{sc}=
	\frac{2}{|\lambda_s|}\({\Delta_d}_m^2-{\Delta_d}_0^2\)
	\\-2\int_{-\mu}^\Lambda\nu(\xi)
	\(\sqrt{\xi^2+{\Delta_d}_m^2}-\sqrt{\xi^2+{\Delta_d}_0^2}\)
	\\+
	2\int^{\sqrt{x^2-{\Delta_d}^2_m}}_{-\sqrt{x^2-{\Delta_d}^2_m}}
	\nu(\xi)
	\sqrt{\xi^2+{\Delta_d}_m^2}
	-
	2x \mathcal{N}(\sqrt{x^2-{\Delta_d}^2_m})
	\leq
	0,
\end{multline} 
 where we distinguish the superconducting gap ${\Delta_d}_m$ in the mixed phase from the superconducting gap ${\Delta_d}_0$ in the pure superconducting phase, and have used a self-consistency equation to reexpress the last term. 
 Since the energy of the pure superconducting phase is minimized by ${\Delta_d}_0$,
 we can conclude that the sum of the first two lines in Eq.~\ref{eq:energydiff} is non-negative. 
 Provided that the density of states does not significantly vary  on the scale of $\sqrt{x^2-{\Delta_d}^2_m}$,
 the third line can be approximated by  
 \begin{multline}
 	2\nu_0{\Delta_d}_m^2\sinh^{-1}\(\frac{\sqrt{x^2-{\Delta_d}_m^2}}{{\Delta_d}_m}\),
 \end{multline} 
where $\nu_0$ is the constant density-of-states, and the full energy difference is positive. 
Hence, at least for a sufficiently smooth density of states,
the antiparallel phase will never be thermodynamically stable. We now analyze the parallel phase.

 \iffalse
 \begin{figure*}[t]
	\includegraphics[width=18cm]{Figures/order_parameters_lxstep01_lsstep002_mu5.png}
	\caption{Order parameter dependence on dimensionless interaction constants $g_s=\nu_{max}\lambda_s$, $g_x=\nu_{max}\lambda_x$ in the parallel phase for $\nu(\epsilon)=\nu_{max}\gamma^2/\(\gamma^2+(\xi-\xi_s)^2\)$with $\xi_s=0$ and $\gamma=1$.	Normalized exciton self-energy $x/\epsilon_F$ is plotted on A). For small absolute values of $g_s$, the exciton condensate appear at the point $g_x=1$ in accordance to the Stoner criterion. Interlayer superconductivity ${\Delta_i}$ is vanishingly since the Fermi energy is tuned to the peak in the density of states. In this calculation, $\Lambda=10 \ meV$, $\epsilon_F=5 \ meV$, $\gamma=1 \ meV$, $\alpha=1$, $\xi_s=0\ meV$. We performed a minimization of energy as a function of the Bogolyubov angles using conjugate graduate method.}
	\label{fig:parallel-phase:van-hove}
\end{figure*}
\fi
\subsection{Parallel phase}
The parallel phase is defined by the conditions $\sin(\psi_+)=0$, $\sin(\psi_-)=0$, so that $\psi_t$ and $\psi_b$ are equal to within a multiple of 
$2\pi$.  This condition, however, leaves an ambiguity since the $\cos(\psi_+)\cos(\psi_-)$ factor 
present in the dispersion relation can be either $+1$ or $-1$. %(on Fig.\ref{fig:parallel-phase:energy-vs-phi} $+1$ and $-1$ correspond to a saddle point and minima)
We distinguish these possibilities by introducing $f=0,1$ such that $\cos(\psi_+)\cos(\psi_-)=(-1)^f$. 
The quasiparticle energies can be expressed as:
\begin{equation}\label{phase-diagram:parallel:dispersion}
	\epsilon_\pm=\sqrt{(\xi \pm x)^2+({\Delta_d} \pm (-1)^f {\Delta_i})^2}.
\end{equation}
We can interpret this phase as two superconductors: one with Cooper pairs formed out of symmetric combinations of layers, another -- from antisymmetric combinations.  Indeed, the mean-field Hamiltonian matrix \eqref{phases:ham-matrix} is block-diagonal in this basis. Note here that even though the transformation to the symmetric and antisymmetric states block-diagonalizes $\mathcal{H}(p)$, energy density $E$ \eqref{phases:total-energy} will still have terms that couple the two unless $|\lambda_s|=\lambda_x$.
Solutions with $f=0/1$ then correspond to phases with in which either symmetric or antisymmetric superconducting gaps are larger.
Variation of the full energy density \eqref{phases:total-energy} with respect to ${\Delta_i}$ yields
\begin{multline}\label{phase-diagram:parallel:energy-derivative:sigma}
	\frac{\delta E}{\delta {\Delta_i}}
	=
	-\frac{4{\Delta_i}}{\lambda_x}
	-
	{\Delta_i} \int 
	\nu(\xi)
	\(
	\frac{1}{\epsilon_-}
	+
	\frac{1}{\epsilon_+}
	\)
	\\+
	(-1)^f
	\Delta_d \int 
	\nu(\xi)
	\(
	\frac{1}{\epsilon_-}
	-
	\frac{1}{\epsilon_+}
	\)
\end{multline}
We see here that the ${\Delta_i}=0$ point is not an energy extremum 
whenever both exciton condensates and superconductivity are present. 
Note that the superconducting gap of the symmetric quasiparticles is $\Delta_d+ (-)^f{\Delta_i}$, 
while for the antisymmetric quasiparticles is $\Delta_d- (-)^f {\Delta_i}$.  Because the exciton condensate breaks the symmetry between quasiparticle bands, 
one should expect that $\Delta_d+{\Delta_i}\neq\Delta_d-{\Delta_i}$. 
A complimentary explanation is that even though the interlayer interaction is repulsive, 
the BCS instability present within each layer together with interlayer coherence 
induces a response in the interlayer Cooper channel: $\Delta_i$ is induced by the coexistence of the exciton condensate and superconductivity. It would not acquire non-zero value in the isolation.

If we write the self-consistency equations in terms of
$\Delta_-\equiv\Delta_d-(-1)^f{\Delta_i}$, $\Delta_+\equiv \Delta_d+(-1)^f{\Delta_i}$ and $x$ we obtain,
\begin{multline}\label{phase-diagram:parallel:delta+}
	\frac{1}{\lambda_-^2-\lambda_+^2}(\lambda_+\Delta_+-\lambda_-\Delta_-)
	\\=
	\frac{\Delta_+}{2}
	\int \frac{d\xi \nu(\xi)}{\sqrt{(\xi+x)^2+\Delta_+^2}},
	\end{multline}
\begin{multline}
	\label{phase-diagram:parallel:delta-}
\frac{1}{\lambda_-^2-\lambda_+^2}(\lambda_+\Delta_--\lambda_-\Delta_+)
	\\=
	\frac{\Delta_-}{2}
	\int \frac{d\xi \nu(\xi)}{\sqrt{(\xi-x)^2+\Delta_-^2}},
\end{multline}
\begin{multline}
	\label{phase-diagram:parallel:x}
		\frac{4}{\lambda_+-\lambda_-}x=\int \frac{d\xi\nu(\xi)(x-\xi)}{\sqrt{(\xi-x)^2+\Delta_-^2}}
	\\+
	\int \frac{d\xi\nu(\xi)(x+\xi)}{\sqrt{(\xi+x)^2+\Delta_+^2}}.
\end{multline}
Here $\lambda_+\equiv(\lambda_s+\lambda_x)/2$ and $\lambda_-\equiv(\lambda_s-\lambda_x)/2$. Note that even though the mean-field 
Hamiltonian matrix is block diagonal in the basis of the symmetric and antisymmetric combinations of two layers, the energy density has a term of the form $\Delta_+\Delta_-$ because $\lambda_s\neq\lambda_x$. 
The mixed state solution summarized in Fig.\ref{fig:parallel-phase:underdoped} exists for $\nu(0)\lambda_x\geq 1$. 
  \begin{figure}[htb!]
	\includegraphics[width=\columnwidth]{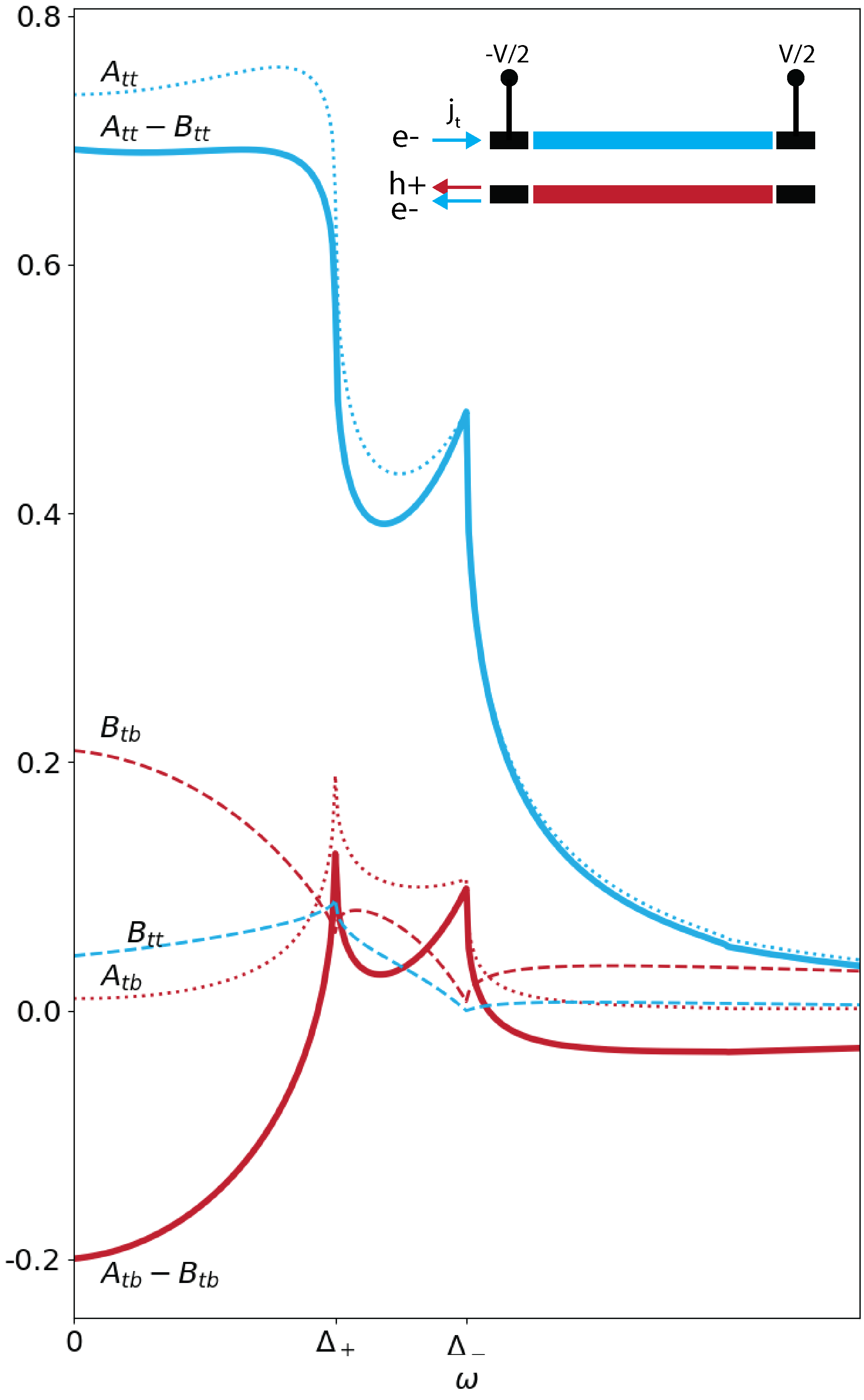}
	\caption{Probabilities for an electron coming from the top lead at energy $\omega$ to be reflected as a hole/electron in the same layer ($A_{tt}$/$B_{tt}$), and as a hole/electron in the opposite layer ($A_{tb}$/$B_{tb}$). 
	The quantity $A_{tb}-B_{tb}$, proportional to the differential conductance in the bottom layer, 
	changes sign as function of $\omega$. At low energies, the conductance is dominated by electron-electron reflection. 
	At energies around superconducting gaps, the conductance is dominated by Andreev reflection processes. 
	The inset shows a schematic experimental device with voltage bias $V$ across the top layer. 
	When mesoscopic effects are neglected, the top layer conductance of the device is $\sigma_{t}(V/2)/2$,
	where $\sigma_{t}(V)$ is the differential conductance between the top left lead and the top layer,
	and the transconductance driven by excitonic order is $\sigma_{b}(V/2)/2$, where 
	$\sigma_{b}(V)$ is the conductance between the top left lead and the bottom layer.}
	\label{fig:exp:setup}
\end{figure}
  \begin{figure}[htb!]
	\includegraphics[width=\columnwidth]{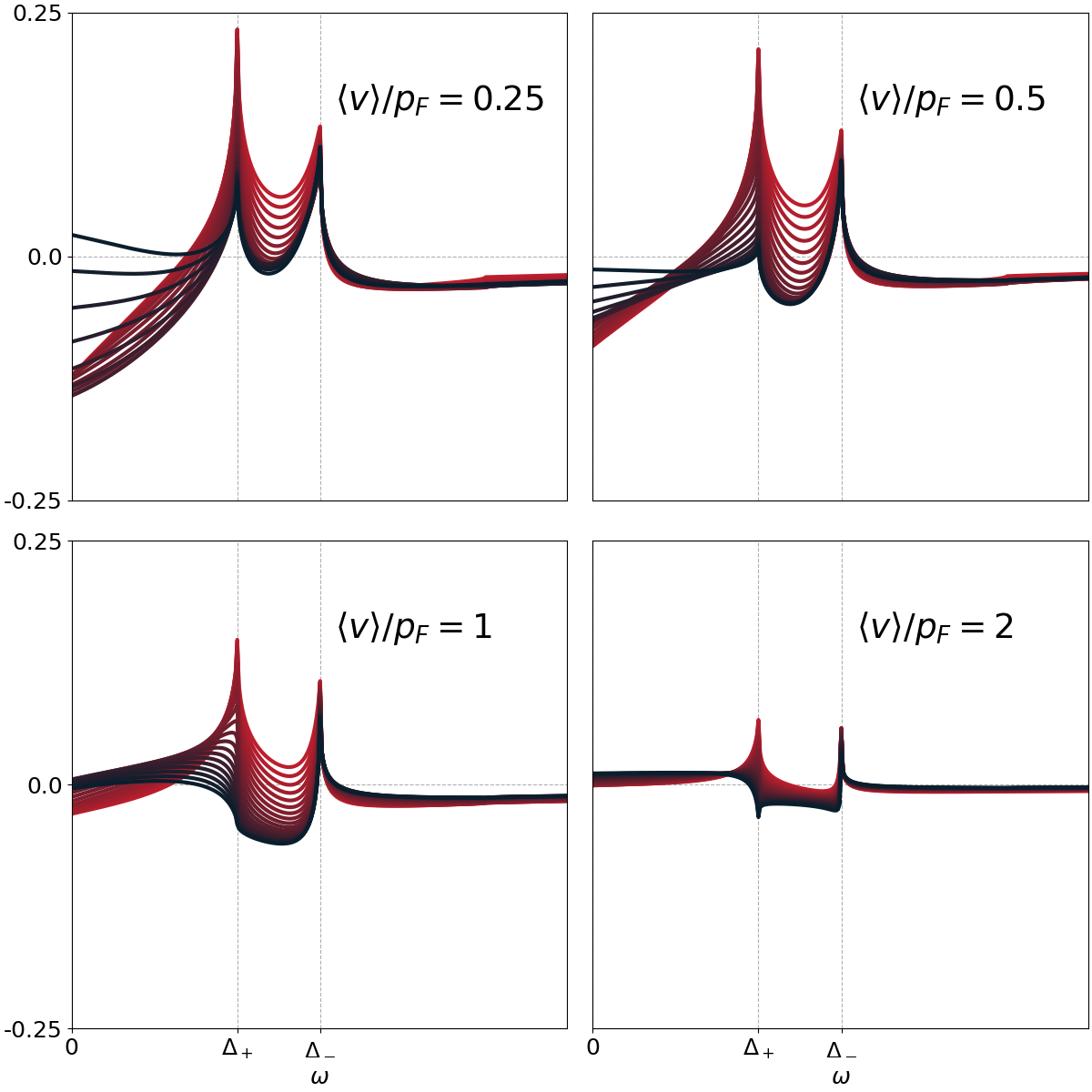}
	\caption{Influence of interface disorder on the difference between hole (Andreev process) and 
	electron interlayer reflection coefficients $A_{tb}-B_{tb}$. 
	Disorder is modeled by a delta-function potential at the interface $v_l\delta(z)$. 
	On each subplot, the ratio of $\<v\>=(v_t+v_b)/2$ to the Fermi momentum is kept fixed. $\Delta v/p_F=(v_t-v_b)/(2p_F)$ is swept from -1.0 (red) to +0.8 (black) with step 0.1.}
	\label{fig:exp:disorder}
\end{figure}
In the numerical calculation, to treat both $f=0$ and $f=1$ minima on the equal footing, we allowed the interlayer superconducting gap $\Delta_i$ to take negative values. This extension does not have any physical significance, as the discreteness of $f$ comes from the energy minimization. Negative value of $\Delta_i$ corresponds to the $f=1$ minimum while positive values will correspond to the $f=0$ minimum. Our calculation shows that $f=1$ minimum tend to be more energetically favorable for larger values of intralayer attraction, which has a simple physical meaning: for $|\lambda_s|\approx \lambda_x$ the symmetric and the antisymmetric superconductors are effectively decoupled and thus larger gap correspond to a larger effective Fermi energy $\mu \pm x$ (see Appendix B).
Along the first order phase boundary in $g_s$/$g_x$ space separating mixed and pure superconducting phases, 
the two states have identical energies.  It follows that the differential energy changes 
$dE_{sc}$ in the superconducting phase and $dE_{mix}$ in the mixed phase should be the same.
It follows that the slope of the phase boundary lines 
\begin{equation}\label{phase-diagram:parallel:coexistence-curve}
	\frac{dg_x}{dg_s}
	=
	\frac{\frac{\d E_{sc}}{\d g_s}-\frac{\d E_{mix}}{\d g_s}}{\frac{\d E_{mix}}{\d g_x}}.
\end{equation}
Partial derivatives in the formula will be proportional to order parameters in each phase: $\d E_{sc,mix}/\d g_s=-2\Delta_{sc,mix}^2/g_s^2$, $\d E_{sc}/\d g_x=2(x^2-{\Delta_i}^2)/g_x^2\approx 2\mu^2/g_x^2$. Formation of the mixed state accompanied by the immediate decrease in the superconducting gap at least for large enough $|g_s|$. Therefore, $dg_x/dg_s<0$ --  for larger attractive interaction $g_s$ transition will happen at a larger critical $g_x$ in agreement with Fig. \ref{fig:parallel-phase:underdoped}. Physically, it corresponds to the fact that the mixed parallel and purely superconducting phases are competing. Because with larger attraction within the layer it is more energetically beneficial to have larger superconducting gaps, the purely superconducting phase becomes could be more energetically favorable and thus the boundary between the phases goes up.
Further increase in the interlayer repulsion leads to the enhancement of the superconductivity.  It is easier to see however that in the limit of infinite repulsion $\lambda_x\gg|\lambda_s|$ between layers the gap equations are \eqref{phase-diagram:parallel:delta+} and \eqref{phase-diagram:parallel:delta-} are again governed by the through the intralayer attraction $\lambda_s$.

\iffalse Assuming that $\Delta_{sc}^2-\Delta_{mix}^2=g_s^2 c$ with $c$ being weakly dependent on $g_s$ and $g_x$ at the interface, by integrating \label{phase-diagram:parallel:coexistence-curve} from point $(g_x,g_s)=(1,0)$ we get $g_x=(1+cg_s/\mu^2)^{-1}$ along the interface.\fi

\section{Non-local Andreev Reflection}

Having established that an exotic mixed state with both excitonic and Cooper pair order can exist,
we now discuss Andreev reflection experiments in separately contacted bilayers that 
can be used to detect their presence. Consider metallic leads separately connected to the top and the bottom 
layers of our system. An electron incident on the interface between the top lead and the top layer
has equal weight in symmetric and antisymmetric quasiparticle 
channels $\ket{t}=(\ket{+}+\ket{-})2^{-1/2}$. Since the parallel state does not couple $\ket{+}$ with $\ket{-}$, we can 
consider their transmission contributions separately, at least if disorder is neglected.
Incoming from the top quasiparticles are reflected with probabilities $A_{tt}$ ($tt$ stands for top$\to$top) and $A_{tb}$ ($tb$ stands for top$\to$bottom) as holes,
and with probabilities $B_{tt}$ and $B_{tb}$ as electrons, to the top and bottom layers respectively. Respective amplitudes are denoted with small letters ($a_{tt}$, $a_{tb}$, $b_{tt}$, $b_{tb}$).
The calculation of these reflection amplitudes is discussed below.

If we apply $V$ voltage across the top part of the interface, the 
top and bottom layer differential conductances will be connected to reflection coefficients in a normal fashion.
The total currents in the top and bottom layers are 
\begin{equation} 
I_{t/b} = e \nu_0 \int_{0}^{eV} d \omega \; \sigma_{t/b}(V)   
\end{equation} 
where 
\begin{eqnarray}
 \sigma_t(V)&\equiv& dj_t/dV=e^2 v_f\nu_0 (1-B_{tt}(V)+A_{tt}(V)), \nonumber \\
 \sigma_b(V)&\equiv& dj_b/dV=e^2 v_f\nu_0 (A_{tb}(V)-B_{bt}(V)),
 \end{eqnarray}
$e$ is an electron charge, and $v_f\nu_0$ is the product of the Fermi velocity and the 
density of states.  Note that the combination 
 $A_{tb}-B_{tb}$ ($A_{tb}$ is a probability to reflect an incoming from the top electron as hole to the bottom layer, $B_{tb}$ is a probability to reflect it as an electron) of cross-layer reflection probabilities can be measured experimentally.
 We derive expressions for the scatterring amplitudes in the Appendix using the Bogoliubov-de-Gennes equations. The amplitude to reflect an electron as a hole in either symmetric or antisymmetric states, at least in the limit $x\ll \mu$, around 1 at $\omega=0$ similarly to a single N-S junction. However, the amplitude to reflect an incoming electron as an electron is no longer negligible even in the $\Delta_d\ll \mu$ case. The electron reflection amplitude at $\omega=0$ is given by $b_\pm\approx\mp x/(2\mu\pm x)$.  
 Because $b_+$ and $b_-$ have opposite signs, the probability to reflect an electron to the bottom layer $B_{tb}=|b_+-b_-|^2/4$ is substantilly higher than the probability of the corresponding Andreev process, which is almost zero. 
For the the same reason electron reflection in the top layer is unimportant for small $\omega$ and instead the Andreev process dominates (Fig.\ref{fig:exp:setup}). 
When energy is equal to the $+$ or $-$ gap and the gap $\Delta_\pm/\mu\ll 1$, 
the corresponding momenta in the leads and device match, giving rise to peaks in Andreev reflection and minima in electron reflection.
As a result the Andreev process is stronger than the electron reflection in the bottom layer. 
Altogether, we expect $A_{tb}-B_{tb}$ to have a $\mu$-like shape, with negative values
around $\omega=0$ and positive peaks at the superconducting gaps (Fig.\ref{fig:exp:setup}).
 
\iffalse $A_{tb}-B_{tb}$ is a quantity one can measure experimentally.
 If we apply $V$ voltage across the top part of the interface, top and bottom differential conductance will be connected to reflection coefficients in a normal fashion: $\sigma_t(V)\equiv dj_t/dV=e^2v_f\nu_0 (1-B_{tt}(V)+A_{tt}(V))$, $\sigma_b(V)\equiv dj_b/dV=e^2v_f \nu_0 (A_{tb}(V)-B_{bt}(V))$, where $e$ is an electron charge, and $v_f\nu_0$ is a characteristic of a lead. As a result, we expect the differential conductance in the bottom layer to change sign several times as function of the applied voltage, with robust positive peaks at values corresponding to the superconducting gaps.
If we model disorder by a delta-function $v_l\delta(z)$ at the interface in each layer, $\<v\>=(v_t+v_b)/2$ will play a role identical to the disorder in a regular superconductor: by damping Andreev reflection at subgap energies and increasing electron-electron reflection in symmetric and antissymetric superconductors separately. A dimensionless quantity controlling its importance is $\<v\>/v_F$, where $v_F$ is the Fermi velocity in the lead. A difference $\Delta v=(v_t-v_b)/2$ will couple different superconductors at the interface. Its importance however, also controlled by $\Delta_d v/v_F$ and for values $\Delta_d v/v_F<0.5$ does not qualitatively change $\mu$-like shape of the differential conductance  (Fig.\ref{fig:exp:disorder}). Larger values of $\Delta_d v$ raise low-energy part of the differential conductance to positive energies. \fi

If we follow the standard practice \cite{blonder1982transition} of modelling
interface disorder by a delta-function potentials $v_l\delta(z)$, $\<v\>=(v_t+v_b)/2$ will play a role identical to that of  the disorder in a regular superconductor
by damping Andreev reflection at subgap energies and increasing electron-electron reflection in both symmetric and antisymetric channels. The dimensionless quantity controlling its importance is $\<v\>/v_F$, where $v_F$ is the Fermi velocity. Any difference $\Delta v=(v_t-v_b)/2$ between layers in the disorder parameter values 
will couple the symmetric and antisymmetric quasiparticle channels. 
The importance of these corrections
is however also controlled by $\Delta_d v/v_F$, and for values $\Delta_d v/v_F<0.5$  there is 
no qualitative change in the $\mu$-like shape of the differential conductance  (Fig.\ref{fig:exp:disorder}). Sufficiently large values of $\Delta_d v/v_F$ raise the low-energy part of the differential conductance to positive values. 

\section{Discussion}

In recent years experimenters have established moir\'e heterojunctions as attractive 
platforms \cite{kennes2021moire} for new types of two-dimensional electron ground states. 
In this Article, we have explored the phase diagram of bilayers with attractive effective
interactions ($\lambda_s$) within each layer and repulsive interactions ($\lambda_x$) between layers.
Two-dimensional bilayers with repulsive interlayer interactions can \cite{eisensteinQHE} have 
excitonic insulator ground states that are counterflow superfluids and have spontaneous 
interlayer phase coherence.  
Our work is motivated by the discovery  of superconductivity, and hence attractive 
effective interactions, in graphene bilayers \cite{cao2018unconventional,yankowitz2019tuning} 
and trilayers \cite{park2021tunable}.
We therefore address the possibility of states
that have both superconductivity and interlayer coherence, and 
study how the occurrence of such exotic states would be manifested by
generalized Andreev effects in separately contacted bilayers.

With this goal we have performed mean-field calculations for a model bilayer Hamiltonian with
two identical layers, and attractive intralayer and repulsive interlayer interactions.
We have neglected tunneling $\tau$ between the layers, assuming that the 
bilayers are separated by dielectric layers. We stress here that unlike 
in \cite{sager2020potential}, we study a system with the ordering between layers alike the magnetism. The difference is that unlike in the scenario with excitonic condensation, there is no gap established directly through the interlayer coherence.

Our calculations show that a mixed phase with both an interlayer exciton coherence and 
superconductivity appears over a wide range of model parameters.
Interlayer coherence couples the two superconducting order parameters with 
energy extrema occurring when the pair amplitudes are in phase (parallel) and out of phase (antiparallel).
We find that the energy of the antiparallel phase is higher than that of the pure 
superconducting phase for any sufficiently smooth density of states. 
In the parallel phase superconducting state, 
interlayer coherence appears as a strong-coupling instability of states 
with superconductivity in each layer, and 
is established when $g_x \equiv \nu(\epsilon_F)\lambda_x>1$. 
In twisted bilayer graphene we estimate that $\lambda_x\approx e^2/(4\pi k_F) \approx 100 \ meV nm^2$, 
and that the density of states is around $\nu(\xi_s)\approx 10^{-2} \ nm^{-2} \ meV^{-1}$ \cite{xie2020weak}, implying that $g_x=1$ is within reach. Since the superconducting transition temperature 
in tBG is $\sim 1 K$, the value we have chosen for $|g_s|\lessapprox 1$, is not unrealistic.
After interlayer coherence is established, further $\lambda_x$ increases 
cause the superconducting order parameters to decay.
Note that self-consistent solution with both exciton order parameter
$x$ and direct intralayer pairing $\Delta_d$ 
nonzero also have non-zero inter-layer pairing $\Delta_i$. 

As a convenience, we have limited our considerations to the case of two completely identical layers. 
This condition is however not important for the stability of the phase. Consider for example the 
case in which the Fermi energies in the two layers are slightly different: $\mu_{t,b}=\mu\pm\Delta\mu$. 
This change induces a perturbation to the mean field Hamiltonian matrix of the 
form $-\tau_z\sigma^z\Delta \mu$, where $\tau^z$ and $\sigma_z$ are the z-Pauli matrices 
acting on layer and Nambu indices correspondingly.  
After transformation to $+/-$ basis the perturbation acquire form $i\tau^y\sigma_z\mu$. 
As a result, there will be no first-order contribution to the quasparticle energies. 
At second order, only matrix elements between positive and negative energy states 
will contribute. The leading order correction to the quasiparticle energies will be small, 
$\sim (\Delta\mu^2/\sqrt{\Delta^2+x^2})$.  We conclude that small deviations from the perfect layer 
symmetric case do not destroy the state, although they will change the mean-field-equation 
solutions quantitatively.

\section{Acknowledgments}
We acknowledge support by  DOE grant DE-FG02-02ER45958. I.V.B. thanks Nemin Wei,  Ajesh Kumar and Rafi Bistritzer for discussions.

\bibliography{xc_sc_coex}

\begin{thebibliography}{59}
\bibitem [{pha()}]{phase-condition-comment}%
  \BibitemOpen
  \href@noop {} {}\bibinfo {note} {The condition written assuming all three
  order-parameters $\Sigma$, $\Delta$ and $x$ are non-zero. If $\Sigma=0$,
  $\psi_+$ can be any.}\BibitemShut {Stop}%
\end{thebibliography}
\bibliographystyle{apsrev4-1}
\begin{widetext} 
\section*{Appendix A: Bogoloyubov-de-Gennes equation for the parallel state}
Because Andreev reflection in symmetric and antisymmetric states cancel each other at zero voltage bias, it is instrumental to consider also electron-electron reflection. It is still true however that Bogoliubov-de-Gennes equation consists of two uncoupled systems of equations for a fully symmetric case. It is then sufficient to solve only one of them:
\begin{align}
	i\d_t f=-\(\frac{1}{2m}\frac{\d^2}{\d z^2}+\mu(z)\)f+\Delta(z)\phi\\
	i\d_t \phi=\(\frac{1}{2m}\frac{\d^2}{\d z^2}+\mu(z)\)\phi+\Delta(z)f
\end{align}
The only difference from the classical \cite{andreev1965thermal,blonder1982transition} setup is that $\mu(z)$ is a function of a coordinate to take into account the presence of the interlayer coherence inside the system.  Fourier transform with respect to the time $f(t)=\int e^{-i\omega t}f_\omega$ will give
\begin{align}
		\frac{1}{2m}\frac{\d^2}{\d z^2}f=-\(\omega+\mu(z)\)f+\Delta(z)\phi\\
	\frac{1}{2m}\frac{\d^2}{\d z^2}\phi=(\omega -\mu(z))\phi-\Delta(z)f
\end{align}
Or, alternatively,
\begin{equation}
	\frac{1}{2m}\frac{\d^2}{\d z^2}
	\begin{pmatrix}
		f\\ \phi
	\end{pmatrix}
	=
	\begin{pmatrix}
		-(\omega+\mu(z)) && \Delta(z) \\
		-\Delta(z) && (\omega-\mu(z))
	\end{pmatrix}
		\begin{pmatrix}
		f\\ \phi
	\end{pmatrix}
\end{equation}
Let's assume that $\Delta$ changes abruptly from 0 to $\Delta_0$ and similarly $\mu$ changes from $\mu$ to $\mu+x$: 
\begin{align}
	\mu(z)=\mu+\theta(z)x\\
	\Delta(z)=\theta(z)\Delta,
\end{align}
where $x$ can be both positive and negative. We solve it by writing it as a system of the first order equations
\begin{align}
	\frac{1}{2m}
	\frac{\d }{\d z}\begin{pmatrix}
		f' \\
		\phi' \\
		f \\
		\phi
	\end{pmatrix}
	=
		\begin{pmatrix}
		0 && 0 && -(\omega+\mu) && \Delta  \\
		 0 && 0 && -\Delta && (\omega-\mu) \\
		1/2m && 0 && 0 && 0 \\
		0 && 1/2m && 0 && 0 
	\end{pmatrix}
	\begin{pmatrix}
		f' \\
		\phi' \\
		f \\
		\phi \\
	\end{pmatrix}
\end{align}
Eigenvalues of this matrix are $\pm \frac{i}{\sqrt{2m}}\(\mu+\sqrt{\omega^2-\Delta^2}\)^{1/2}$, $\pm \frac{i}{\sqrt{2m}}\(\mu-\sqrt{\omega^2-\Delta^2}\)^{1/2}$. %Among them, for $\omega<\Delta$ only $\frac{i}{\sqrt{2m}}\(\mu+\sqrt{\omega^2-\Delta^2}\)^{1/2}$ will be both decaying in $z\to\infty$ limit and right-propagating. It seems however that for consistency I also have to include left-propagating solution $\frac{i}{\sqrt{2m}}\(\mu-\sqrt{\omega^2-\Delta^2}\)^{1/2}$.
A general solution for $\omega<\Delta$ will be
\begin{equation}
\begin{pmatrix}
			f \\
		\phi
	\end{pmatrix}
	=
	\frac{C}{\sqrt{2}}e^{iz\sqrt{2m}\(\mu+i\sqrt{\Delta^2-\omega^2}\)^{1/2}}
	\begin{pmatrix}
			\sqrt{\frac{\omega+i\sqrt{\Delta^2-\omega^2}}{\Delta}} \\
			\sqrt{\frac{\omega-i\sqrt{\Delta^2-\omega^2}}{\Delta}} 
	\end{pmatrix}
	+
		\frac{D}{\sqrt{2}}e^{-iz\sqrt{2m}\(\mu-i\sqrt{\Delta^2-\omega^2}\)^{1/2}}
	\begin{pmatrix}
			\sqrt{\frac{\omega-i\sqrt{\Delta^2-\omega^2}}{\Delta}} \\
			\sqrt{\frac{\omega+i\sqrt{\Delta^2-\omega^2}}{\Delta}} 
	\end{pmatrix}
\end{equation}
I will denote in what follows $\sqrt{\frac{\omega+i\sqrt{\Delta^2-\omega^2}}{\Delta}} =u$ and $\sqrt{\frac{\omega-i\sqrt{\Delta^2-\omega^2}}{\Delta}}=v$ and $k_{sC}=\sqrt{2m}\(\mu+i\sqrt{\Delta^2-\omega^2}\)^{1/2}$, $k_{sD}=\sqrt{2m}\(\mu-i\sqrt{\Delta^2-\omega^2}\)^{1/2}$.
On the left hand side equations for a hole and an electron decouple and as a solution I get instead:
\begin{equation}
	\begin{pmatrix}
			f \\
		\phi
	\end{pmatrix}
	=
	e^{i\sqrt{2m}z\sqrt{\omega+\mu_N}} 
	\begin{pmatrix}1 \\ 0\end{pmatrix}
	+
	be^{-i\sqrt{2m}z\sqrt{\omega+\mu_N}} 
	\begin{pmatrix}1 \\ 0\end{pmatrix}
	+
	a\begin{pmatrix}0 \\ 1\end{pmatrix}
	e^{i\sqrt{2m}z\sqrt{\mu_N-\omega}} 
\end{equation}
The last piece is the reflected hole. Now boundary conditions read:
\begin{align}
	1+b=\frac{C}{\sqrt{2}}u+\frac{D}{\sqrt{2}}v\\
	a=\frac{C}{\sqrt{2}}v+\frac{D}{\sqrt{2}}u\\
	k_e(1-b)=\frac{C}{\sqrt{2}}uk_{sC}-\frac{D k_{sD}}{\sqrt{2}}v\\
	k_h a=\frac{C}{\sqrt{2}}vk_{sC}-\frac{D k_{sD}}{\sqrt{2}}u,
\end{align}
from which we conclude that 
\begin{align}
2=\frac{C u}{\sqrt{2}}\(1+\frac{k_{sC}}{k_e}\)+\frac{Dv}{\sqrt{2}}\(1-\frac{k_{sD}}{k_e}\)\\
0=\frac{C v}{\sqrt{2}}
\(1-\frac{k_{sC}}{k_h}\)
+
\frac{D u}{\sqrt{2}}
\(1+\frac{k_{sD}}{k_h}\)
\end{align}
hence
\begin{equation}
	\frac{D}{C}=-\frac{v}{u}\frac{1-\frac{k_{sC}}{k_h}}{1+\frac{k_{sD}}{k_h}}, 
\end{equation}
and so
\begin{equation}
	2=\frac{C u}{\sqrt{2}}\(1+\frac{k_{sC}}{k_e}\)-\frac{C v^2}{u\sqrt{2}}\(1-\frac{k_{sD}}{k_e}\)\frac{1-\frac{k_{sC}}{k_h}}{1+\frac{k_{sD}}{k_h}}
\end{equation}
consequently
\begin{equation}
	C=\frac{
	2^{3/2}u\(1+\frac{k_{sD}}{k_h}\)
	}{
	u^2\(1+\frac{k_{sC}}{k_e}\)\(1+\frac{k_{sD}}{k_h}\)-v^2\(1-\frac{k_{sD}}{k_e}\)\(1-\frac{k_{sC}}{k_h}\)}
\end{equation}
\begin{equation}
	D=-\frac{2^{3/2}v\(1-\frac{k_{sC}}{k_h}\)}
	{u^2\(1+\frac{k_{sC}}{k_e}\)\(1+\frac{k_{sD}}{k_h}\)-v^2\(1-\frac{k_{sD}}{k_h}\)\(1-\frac{k_{sC}}{k_h}\)}
\end{equation}
Expressions for amplitudes $a$ and $b$:
\begin{equation}\label{app1:a}
	a=\frac{
	2uv\(\frac{k_{sD}}{k_h}+\frac{k_{sC}}{k_h}\)
	}{
	u^2\(1+\frac{k_{sC}}{k_e}\)\(1+\frac{k_{sD}}{k_h}\)-v^2\(1-\frac{k_{sD}}{k_e}\)\(1-\frac{k_{sC}}{k_h}\)}
\end{equation}
\begin{equation}\label{app1:b}
	b=
	\frac{u^2\(1+\frac{k_{sD}}{k_h}\)\(1-\frac{k_{sC}}{k_e}\)-v^2\(1-\frac{k_{sC}}{k_h}\)\(1+\frac{k_{sD}}{k_e}\)}{u^2\(1+\frac{k_{sC}}{k_e}\)\(1+\frac{k_{sC}}{k_h}\)-v^2\(1-\frac{k_{sD}}{k_e}\)\(1-\frac{k_{sC}}{k_h}\)}
\end{equation}

It is clear that a drastic change of the potential on the interface will lead to an increase in electron-electron scattering. Let us explore the low energy properties of both $a$ and $b$. Because at $\omega=0$: $k_e=k_h=p_F$, where $p_F$ is a Fermi momentum inside the metallic lead, $u=e^{i\pi/4},v=e^{i3\pi/4} $. As a result, at $\omega=0$:
\begin{equation}
    a=-\frac{i\sqrt{\mu_N}(\sqrt{\mu-i\Delta}+\sqrt{\mu-i\Delta})}{\mu_N+\sqrt{\mu+i\Delta}\sqrt{\mu-i\Delta}}
\end{equation}
\begin{equation}
    b=\frac{(\sqrt{\mu_N}-\sqrt{i\Delta+\mu})(\sqrt{\mu_N}+\sqrt{\mu-i\Delta})}{\mu_N+\sqrt{\mu+i\Delta}\sqrt{\mu-i\Delta}}
\end{equation}
Corresponding probabilities $A_{tt}=|a_++a_-|^2$, $A_{tb}=|a_+-a_-|^2$, $B_{tt}=|b_++b-|^2$, $B_{tb}=|b_+-b-|^2$. Because $a$ for symmetric and antisymmetric states are approximately equal to each other whenever $\mu\gg\Delta$, their difference is typically small if this condition is satisfied.  Surprisingly, electron-electron scattering amplitude do not vanish in the $\frac{\Delta}{\mu_N}\to 0$ limit, because the effective Fermi energies inside and outside the system are different. When the $\omega$ is close to the gap, $u\approx v$, $k_C\approx k_D$, and if the gap is much smaller than $\mu_N$, $k_e\approx k_h$, therefore $b\approx 0$, $a\approx 1$. Thus, we expect the Andreev process to be dominant at frequencies corresponding to one of the gaps. If the disorder is present on the interface in the form $V_l(z)=v_l\delta(z)$, Bogoliubov-de-Gennes equations are still decoupled in the bulk, but amplitudes are now coupled through the boundary conditions:
\begin{align}\label{app:andreev:disorder:continuity}
	1+b_{\pm}=\frac{C_{\pm}}{\sqrt{2}}u_{\pm}+\frac{D_{\pm}}{\sqrt{2}}v_{\pm}\\
	a_{\pm}=\frac{C_{\pm}}{\sqrt{2}}v_{\pm}+\frac{D_{\pm}}{\sqrt{2}}u_{\pm}
\end{align}
\begin{align*}\label{app:andreev:disorder:differentiability}
	\frac{i}{2m}\(u_{\pm} 2^{-1/2}C_\pm k_{\pm sC}-v_{\pm} 2^{-1/2}D_\pm k_{\pm sD}-k_e+k_e b_{+/-}\)
	=
	\<v\>(C_{\pm}u_{\pm}2^{-1/2}+D_{\pm}v_{\pm}2^{-1/2})-\Delta v(C_\mp u_{\mp}2^{-1/2}+D_\mp v_{\mp}2^{-1/2})
	\\
	\frac{i}{2m}\(v_{\pm}C_\pm 2^{-1/2} k_{\pm sC}-u_{\pm}D_\pm 2^{-1/2} k_{\pm sD}-k_h a_{+/-}\)
	=
	\<v\>(C_{\pm}v_{\pm}2^{-1/2}+D_{\pm}u_{\pm}2^{-1/2})-\Delta v(C_\mp v_{\mp}2^{-1/2}+D_{\mp}u_{\mp}2^{-1/2}),
\end{align*}
where $\Delta v=(v_t-v_b)/2$, $\<v\>=(v_t+v_b)/2$.

\section*{Appendix B: Green's function formalism}
In this appendix we present equations of motion for zero-temperature Green's function and derive an important symmetry of self-energy in a way different from the one we have taken in the main text.
In addition to a normal Green's function, introduce a set of additional functions:
\begin{align}
	i {F^l}^\dagger_{\sigma\sigma'}(p,t_1-t_2)
	=
	\<T\(c^\dagger_{pl\sigma}(t_1)c^\dagger_{-pl\sigma'}(t_2)\)\>e^{-i2\mu_l t_1}\\
	i {F^l}_{\sigma\sigma'}(p,t_1-t_2)
	=
	\<T\(c_{pl\sigma}(t_1)c_{-pl\sigma'}(t_2)\)\>e^{i2\mu_l t_1}\\
	i {H^{tb}}_{\sigma}(p,t_1-t_2)
	=
	\<T\(c_{pt\sigma}(t_1)c^\dagger_{pb\sigma}(t_2)\)\>e^{i(\mu_t-\mu_b) t_1}\\
	i {D^{tb}}^\dagger_{\sigma\sigma'}(p,t_1-t_2)
	=
	\<T\(c^\dagger_{pt\sigma}(t_1)c^\dagger_{-pb\sigma'}(t_2)\)\>e^{-i(\mu_t+\mu_b) t_1}\\
		i {D^{tb}}_{\sigma\sigma'}(p,t_1-t_2)
	=
	\<T\(c_{pt\sigma}(t_1)c_{-pb\sigma'}(t_2)\)\>e^{i(\mu_t+\mu_b) t_1}
\end{align}
Averages are taken over ground states with fixed and, in general, different number of particles of each kind. Because of this, averages are no longer dependent on the time difference. To mitigate this problem, averages are multiplied by an exponential with the difference between energies on the left and the right. I will only consider spatially uniform solutions and without spontaneous magnetization. 
\\First, look at the equation of motion for the normal and the anomalous Green's functions:
\begin{align} \label{eom:G}
	\(i\frac{\d }{\d t}-\epsilon_{pt}\)G_{t\uparrow}(p,t)
	=
	{F^t}^\dagger_{\downarrow\uparrow}(p,t)\Delta_{\uparrow\downarrow}^{tt}
	+
	{D^{bt}}^\dagger_{\downarrow\uparrow}(p,t)\Delta^{tb}_{\uparrow\downarrow}
	+
	H^{bt}_{\uparrow}(p,t)X^{tb}_{\uparrow}
	+\delta(t)
	\\\label{eom:F}
	\(i\frac{\d }{\d t}+\epsilon_{pt}-2\mu_t\) {F^t}^\dagger_{\downarrow\uparrow}(p,t)
	=
	\bar{\Delta}_{\uparrow\downarrow}^{tt}G_{t\uparrow}(-p,t)
	+
	H^{bt}_\uparrow(-p,t)\bar{\Delta}^{bt}_{\uparrow\downarrow}
	-
	{D^{bt}}^\dagger_{\downarrow\uparrow}(p,t)
	X_\downarrow^{bt},
\end{align}
where I defined self-energies:
\begin{align}\label{gap:delta}
	\Delta^{ll}_{\sigma\sigma'}=-\frac{i}{S}\sum V_{ll}(q)F^l_{\sigma\sigma'}(-p-q,0+)\\
	\label{gap:x}
		X^{ll'}_{\sigma}=\frac{i}{S}\sum V_{ll'}(q)H^{ll'}_{\sigma\sigma'}(p-q,0+)\\
		\label{gap:sigma}
				\Delta^{ll'}_{\sigma\sigma'}=-\frac{i}{S}\sum V_{ll'}(q)D^{ll'}_{\sigma\sigma'}(-p+q,0+),
\end{align}
where $l\neq l'$, $S$ is the area, $V_{ll}(V_{ll'})$ is the intralayer (interlayer) interaction. Note here that positive $V_{ll}$ means repulsion within the layer and similarly for $V_{ll'}$. The other two equations are:
\begin{align}\label{eom:H}
	\(i\frac{\d }{\d t}-\epsilon_{pb}-(\mu_t-\mu_b)\)H^{bt}_{\uparrow}(p,t)
	=
	\Delta^{bb}_{\uparrow\downarrow}{D^\dagger}^{bt}_{\downarrow\uparrow}(-p,t)
	+
	X^{bt}_\uparrow G^t_\uparrow(p,t)
	+
	\Delta^{bt}_{\uparrow\downarrow}{F^\dagger}^t_{\downarrow\uparrow}(-p,t)\\
	\label{eom:D}
	\(i\frac{\d }{\d t}+\epsilon_{pb}-(\mu_t+\mu_b)\){D^\dagger}^{bt}_{\downarrow\uparrow}(p,t)
	=
	H^{bt}_{\uparrow}(p,t)\bar{\Delta}^{bb}_{\uparrow\downarrow}
	+
	G^t_\downarrow(-p,t)\bar{\Delta}^{tb}_{\uparrow\downarrow}
	-
	{F^\dagger}^t_{\downarrow\uparrow}(p,t)X^{tb}_\downarrow
\end{align}
These equations can be either derived through equations of motion for operators, or diagrammatically.
Equations \eqref{eom:F}-\eqref{eom:D} constrain the form of the mean-field hamiltonian. Indeed, consider a case without spin symmetry breaking: $X^{tb}_\uparrow e^{-i \psi_\uparrow}=X^{tb}_\downarrow e^{-i \psi_\downarrow}$, $\Delta^{tt}_{\uparrow\downarrow}=-\Delta^{tt}_{\downarrow\uparrow}$, $G_\uparrow=G_\downarrow$.
The latter means that each term in \eqref{eom:G} must be invariant under spin flip: \begin{align}
	H^{bt}_\uparrow e^{ i \psi_\uparrow}=H^{bt}_\downarrow e^{ i \psi_\downarrow}\\
		{D^{bt}}^\dagger_{\downarrow\uparrow}(p,t)\Delta^{tb}_{\uparrow\downarrow}
	=
	{D^{bt}}^\dagger_{\uparrow\downarrow}(p,t)\Delta^{tb}_{\downarrow\uparrow}\\
	{F^t}^\dagger_{\downarrow\uparrow}(p,t)=-{F^t}^\dagger_{\uparrow\downarrow}(p,t)
\end{align}
From \eqref{eom:F} it follows then that
\begin{align}
	\frac{\bar{\Delta}^{bt}_{\uparrow\downarrow}}{\bar{\Delta}^{bt}_{\downarrow\uparrow}}
	=
	-
	\frac{H^{bt}_\downarrow(-p,t)}{H^{bt}_\uparrow(-p,t)}
	=
	e^{i(\psi_\uparrow-\psi_\downarrow+\pi)}
\end{align}
Finally, since $\Delta^{tb}_{\uparrow\downarrow}\propto\<c_{t\uparrow}c_{b\downarrow}\>$,  $\Delta^{tb}_{\uparrow\downarrow}=-\Delta^{bt}_{\downarrow\uparrow}$:

\begin{align}\label{symmetries:sigma}
	\frac{\bar{\Delta}^{bt}_{\uparrow\downarrow}}{\bar{\Delta}^{tb}_{\uparrow\downarrow}}
	=
	e^{i(\psi_\uparrow-\psi_\downarrow)}
\end{align}
Note that we did not imply anything about layer symmetry. 

\section*{Appendix C: Stability of the parallel phase}
To explore the stability of the parallel state we first represent the original Hamiltonian in the basis of the symmetric/anti-symmetric states 
\begin{align}
	c_{+,\sigma}=\frac{1}{\sqrt{2}}\(c_{t,\sigma}+c_{b,\sigma}\),\\
	c_{-,\sigma}=\frac{1}{\sqrt{2}}\(c_{t,\sigma}-c_{b,\sigma}\)
\end{align}
and perform the Bogolyubov transformation:
\begin{align}
		c_{p\alpha\downarrow}=u^*_{\alpha p}\gamma_{\alpha-}(p)-v_{\alpha p} \gamma^\dagger_{\alpha+}(-p),\\
	c_{p\alpha\uparrow}=v_{\alpha p}\gamma^\dagger_{\alpha-}(-p)+u^*_{\alpha p} \gamma_{\alpha+}(p).
\end{align}
Since we perform Bogoliubov rotation within symmetric and antisymmetric subspaces separately, there are only two parameters at each k-point. Condition for bogolyubons to be fermions is $|u_{\alpha k}|^2+|v_{\alpha k}|^2=1$. Rewrite energy in terms of angles $u_{\alpha k}=\cos(\theta_{\alpha k})$ and $v_{\alpha k}=\sin(\theta_{\alpha k})$:
\begin{multline}
	E=2\sum\xi_k \sin(\theta_{\alpha k})^2
	-\frac{\lambda_x}{2^3 S}\(\sum(\cos(2\theta_{+ k})-\cos(2\theta_{- k}))\)^2
	+
	\frac{\lambda_+}{2^2 S}\sum \sin(2\theta_{\alpha k})\sin(2\theta_{\alpha k'})
	+
	\frac{\lambda_-}{2S}\sum \sin(2\theta_{+ k})\sin(2\theta_{- k'}),
\end{multline}
where $\lambda_+=(\lambda_x+\lambda_s)/2$, $\lambda_-=(\lambda_s-\lambda_x)/2$. 
Self-energies become
\begin{equation}
	\Delta_d
	=
	-\frac{\lambda_s}{S}\sum_{k'}  \<c_{t\uparrow}(k')c_{t\downarrow}(-k')\>
	\\=
		-\frac{\lambda_s }{2S}\sum_{k'\alpha} u^*_{\alpha k'}v_{\alpha k'}
		=
			-\frac{\lambda_s}{4S}\sum_{k'\alpha}  \sin(2\theta_{\alpha k'})
		\end{equation}
\begin{equation}
	\Delta_i
	=
	-\frac{\lambda_x}{S}\sum_{k'}  \<c_{t\uparrow}(k')c_{b\downarrow}(-k')\>
	=
	-\frac{1}{2S}\sum_{k'} \lambda_x u^*_{\alpha k'}v_{\beta k'}\tau_{\alpha\beta}^z
	=
		-\frac{\lambda_x }{4S}\sum_{k'} \(\sin(2\theta_{+k})-\sin(2\theta_{-k})\)
\end{equation}

\begin{equation}
	x
	=
	\frac{\lambda_x}{S}\sum_{k'} \<c^\dagger_{b\uparrow}(k')c_{t\uparrow}(k')\>
	\\=-\frac{\lambda_x}{4S}\sum_{k} \(\cos(2\theta_{+k})-\cos(2\theta_{-k})\)
\end{equation}
Minimization of energy with respect to the Bogolyubov angles gives
\begin{equation}\label{app:B:angle}
 \tan(2\theta_{k\pm})=\frac{\lambda_+ S_\pm+\lambda_{-}S_{\mp}}{-\xi_{k}\pm x},
\end{equation}
where $d_\pm=\frac{1}{2S}\sum_k \sin(2\theta_{\pm k})$ is the superconducting amplitude. Non-diagonal entries of the matrix of second-order derivatives vanish in the thermodynamic limit. Diagonal entries are 
\begin{align}
    \frac{1}{2}\frac{\d^2 E}{\d \theta_{\pm k}^2}
    =
    \cos(2\theta_{\pm})
    \((-\xi_k\pm x)+\tan(2\theta_{\pm})(d_\pm\lambda_++\lambda_-d_\mp)\)
\end{align}
At the energy extrema from \eqref{app:B:angle}
\begin{align}
    \frac{1}{2}\frac{\d^2 E}{\d \theta_{\pm k}^2}
    =
   \frac{\cos(2\theta_{\pm k})}{-\xi_k\pm  x} \(
   (\xi_k\pm x)^2+
    (d_\pm\lambda_++\lambda_-d_\mp)^2
    \)
\end{align}
The stability requires that $\frac{\d^2 E}{\d \theta_{\pm k}^2}\geq0$ at any $k$-point. Using \eqref{app:B:angle} to express the $\cos(2\theta)$, I get
\begin{equation}
    \frac{1}{2}\frac{\d^2 E}{\d \theta_{\pm k}^2}
    =
    \frac{1}{\sqrt{(\lambda_+d_\pm+\lambda_+d_\mp)^2+(\xi_k\mp x)^2}}\(
   (\xi_k\mp x)^2+
    (d_\pm\lambda_++\lambda_-d_\mp)^2
    \)>0,
\end{equation}
we then conclude that both solutions ($f=1$,$f=0$) will be stable in the thermodynamic limit. Now we use the expression for energy to see if one of the solutions has lower energy than the other. 
\begin{multline}
	E=const-\sum_k \frac{(\xi_k-x)\xi_k}{\sqrt{(d_+\lambda_++\lambda_-d_-)^2+(\xi_k-x)^2}}
	-\sum_k \frac{(\xi_k+x)\xi_k}{\sqrt{(d_-\lambda_++\lambda_-d_+)^2+(\xi_k+x)^2}}
	-\frac{2x^2}{\lambda_x }
	+
	\frac{2 \Delta_d^2}{\lambda_s}
	+
	\frac{2 \Delta_i^2}{\lambda_x}
\end{multline}
Let us isolate a part sensitive to a permutation of $\Delta_+\equiv\Delta_d+\Delta_i$ and $\Delta_-\equiv\Delta_d-\Delta_i$. It reads
\begin{equation}\label{app:B:energy:asym}
    E_{asym}=x\nu\(\int d\xi
    \frac{\xi-x}{\sqrt{\Delta_-^2+(\xi-x)^2}}
	-\frac{\xi+x}{\sqrt{\Delta_+^2+(\xi+x)^2}}\)
\end{equation}
Most of the time we have fully polarized solution $x=\mu$ and thus the contribution from the second term is always negative. The contribution from the first term is only negative for the range energies between $-\mu$ and $\mu$. If the high-energy energy cutoff $\Lambda>2\mu$ we better have $\Delta_+<\Delta_-$ to minimize the energy. We then conclude that $\Delta_i<0$ or, another words, the $f=1$ solution has lower energy in accordance to our numerical calculations. We also see that, because the negative contribution from the first term is proportional to the $-\mu\nu\sqrt{\Delta^2_-+(2\mu)^2}$, there could be be an energy benefit from having superconducting gaps larger than the one predicted through the Macmillan in the absence of the interlayer coherence. Additionally, let us clarify here the dependence of gaps on interaction parameters that follow from the gap equations \eqref{phase-diagram:parallel:delta+} and \eqref{phase-diagram:parallel:delta-}. First, exciton order parameter $x$ acquires constant value $\mu$ almost immediately after the phase transition and that is why we will it ignore it functional dependence deep into mixed phase. If $\lambda_x\gg|\lambda_s|$, 
\begin{equation}
	-\frac{1}{\lambda_s}(\Delta_++\Delta_-)
	\approx
	\frac{\Delta_+}{2}
	\int \frac{d\xi \nu(\xi)}{\sqrt{(\xi+x)^2+\Delta_+^2}},
	\end{equation}
\begin{equation}
-\frac{1}{\lambda_s}(\Delta_-+\Delta_+)
	\approx
	\frac{\Delta_-}{2}
	\int \frac{d\xi \nu(\xi)}{\sqrt{(\xi-x)^2+\Delta_-^2}},
\end{equation}
Then for small $\Delta_-$, $\Delta_+$ it follows that $\Delta_+\approx \Delta_-$ and thus $\Delta_i\approx 0$. Then it immediately follows that $\Delta_d$ does not depend on interaction between layers $\lambda_x$. Look now at the opposite limit $|\lambda_s|\gg\lambda_x$. In this case,  
\begin{equation}
	-\frac{1}{\lambda_x}(\Delta_+-\Delta_-)
	\approx
	\frac{\Delta_+}{2}
	\int \frac{d\xi \nu(\xi)}{\sqrt{(\xi+x)^2+\Delta_+^2}},
	\end{equation}
\begin{equation}
-\frac{1}{\lambda_x}(\Delta_--\Delta_+)
	\approx
	\frac{\Delta_-}{2}
	\int \frac{d\xi \nu(\xi)}{\sqrt{(\xi-x)^2+\Delta_-^2}},
\end{equation}
from which we conclude that gaps depend only on $g_x$ in this limit. Interestingly, such solution would require $\Delta_d<\Delta_i$, which is nonphysical. In the intermediate regime $\lambda_x\approx |\lambda_s|$ we expect the gaps depend on the combination $\lambda_x-\lambda_s$, so that the gaps should be constant along the lines $g_s=g_x+C$.

\end{widetext} 

\end{document}